\documentclass{aa}
\usepackage{graphicx}
\usepackage{psfig}

\begin{document}

\title{Accounting for the anisoplanatic point spread function 
in deep wide-field adaptive optics images 
      \thanks{Based on observations collected at the European Southern
        Observatory, Chile under programs 70.B-0649, 71.A-0482 and 073.A-0603} 
}
\author{}
\author{G. Cresci \inst{1,2} \and R.~I. Davies \inst{2} 
  \and A.~J. Baker \inst{3,4} \and M.~D. Lehnert \inst{2}
  }
\institute{}
\institute{Dipartimento di Astronomia, Universit\'a di Firenze,
    Largo E. Fermi 5, I-50125, Firenze, Italy
        \and Max-Planck-Institut f\"ur extraterrestrische Physik,
        Postfach 1312, D-85741 Garching, Germany        
        \and Jansky Fellow, National Radio Astronomy Observatory
        \and Department of Astronomy, University of Maryland, College 
        Park, MD 20742-2421, United States
	}

\offprints{G. Cresci \\
    \email{gcresci@arcetri.astro.it}}

\date{Received / Accepted}

\abstract{In this paper we present the approach we have used to determine and 
account for 
the anisoplanatic point spread function (PSF) in deep adaptive optics (AO) images 
for the Survey of a Wide Area with NACO (SWAN) at the ESO VLT. 
The survey comprises adaptive optics observations in the
$K_\mathrm{s}$ band totaling $\sim 30\,\mathrm{arcmin}^2$, 
assembled from 42 discrete fields centered on different 
bright stars suitable for AO guiding.
We develop a parametric model of the PSF variations across the field of view  
in order to build an accurate model PSF for every galaxy detected in each of 
the fields. 
We show that this approach is particularly convenient, as it uses only easily 
available data and makes no uncertain assumptions about the stability of the 
isoplanatic angle during any given night. 
The model was tested using simulated galaxy profiles to check its performance 
in terms of recovering the correct morphological parameters; we find that the 
results are reliable up to $K_\mathrm{s} \sim 20.5$
($K_{\mathrm{AB}}\sim22.3$) in a typical SWAN field. 
Finally, the model obtained was used to derive the first results from five 
SWAN fields, and to obtain the AO morphology of 55 galaxies brighter than 
$K_\mathrm{s} = 20$. These preliminary results demonstrate the unique power of 
AO observations to derive the details of faint galaxy morphologies and to 
study galaxy evolution.

        \keywords{instrumentation: adaptive optics -- galaxies: fundamental 
        parameters -- galaxies: statistics -- infrared galaxies}}

\titlerunning{Accounting for the anisoplanatic PSF in deep 
AO images}

\maketitle


\section{Introduction}

In recent years an increasing number of adaptive optics 
(AO) systems have become available, which are capable of obtaining
near-infrared images at the diffraction limits of 8-meter class
telescopes.
The scientific potential of such instrumentation is evident, but careful
analysis of the resulting data is necessary to take account of the
effects introduced by the method and the limitations of wavefront sensing
and correction.
A key point in the detailed analysis of AO images is the determination
of the point spread function (PSF) across the whole field of view. 
However, this task is made more difficult because the performance and
correction of the AO system are strongly anisoplanatic, and depend on
many parameters such as the brightness of the reference guide star and
the structure (height, size-scale, velocity) of the often quickly
changing atmospheric turbulence.
As a result, the PSF produced by an AO system can change rapidly in
both time and position on the frame. 
For natural guide star (NGS) wavefront sensing, an extragalactic science 
target will typically be offset from the guide star by $\Delta \theta 
\sim$ 10\arcsec--30\arcsec\ at best. As a result,  
the on-axis PSF no longer provides a suitable
reference model for the off-axis PSF at the position 
of the scientific target.
Until the advent of multi-conjugate adaptive optics systems on 8-m
class telescopes (e.g., Le~Louarn et al. \cite{lelouarn}) full 
correction over a wide (1--2\arcmin) field will not be possible, and even 
then there will still be some PSF variation across the field 
(e.g., V\'erinaud et al. \cite{verinaud}).
Therefore, a simple and generally applicable technique to model the
variations in the PSF across a given science field would be of benefit
to all wide-field adaptive optics data.

In this paper we present the approach we have used to account for the
anisoplanatic PSF in deep adaptive optics images for the Survey of a Wide Area 
with NACO (SWAN).
In the following section SWAN will be briefly described and motivated. 
In Sect.~\ref{methods} some different proposed methods for PSF
reconstruction will be analyzed, while in 
Sect.~\ref{our} our method will be explained and discussed. 
Computer simulations were used to validate the model, and their
results are shown in Sect.~\ref{simul}. 
The first application of the model to five of the deep science fields
will be presented in Sect.~\ref{5results}, and our 
conclusions follow in Sect.~\ref{concl}. 
  
\section{SWAN} \label{SWAN}

In order to make full use of the new generation of telescopes, it is
necessary to overcome the blurring effects of the atmosphere through
the use of AO systems.  These can allow ground-based telescopes to operate 
at or near the diffraction limit in the near infrared ($\sim 0.06\arcsec$ 
in K-band for an 8 meters telescope), resulting in high 
angular resolution and a low background in each pixel.  In principle, 
such capabilities should offer important benefits for studying how galaxies 
form and evolve in the early universe.  In practice, however, reaping the 
expected rewards has proved difficult, due to the small number of 
extragalactic sources known lying at distances $\Delta \theta \la 30\arcsec$ 
from bright ($V \la 13$) stars suitable for AO guiding.

The prospects for AO cosmology will undoubtedly improve with the 
widespread adoption of laser guide star (LGS) systems, since these impose less 
stringent requirements on the brightness and nearness of stars used for 
tip-tilt correction.  In the mean time, to overcome the current shortage of 
extragalactic AO targets for NGS wavefront sensing, we have undertaken a 
program to identify and characterize faint field galaxies lying close to 
bright, blue stars at high Galactic latitudes (see, e.g., Larkin et al. 
\cite{larkin99}; Davies et al. \cite{davies01}).  Our own 42 southern bright 
star fields were initially imaged at seeing-limited resolution in 
$K_\mathrm{s}$ with SOFI at the ESO {\em New Technology Telescope} (Baker et al. 
\cite{baker}), were followed up with optical imaging (Davies et al. 
\cite{davies05}), and are now targets for VIMOS optical spectroscopy at the 
ESO Very Large Telescope (VLT).
\begin{figure}
    \begin{center}
     \resizebox{\hsize}{!}{\includegraphics[clip]{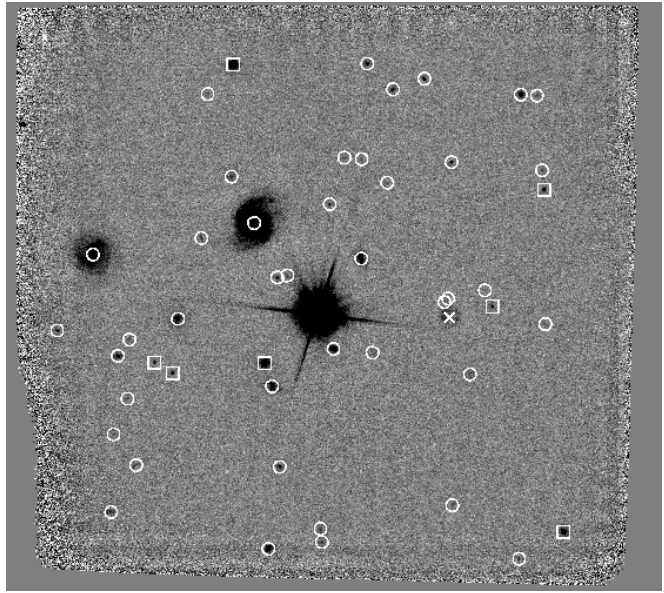}}   
    \caption{Example of a SWAN field: SBSF\,18. The circles are the extended 
    objects detected by SExtractor (SExtractor stellarity index 
    $\mathrm{SSI} < 0.9$), while the squares are point sources ($\mathrm{SSI} 
    \geq 0.9$).  A ghost of the bright guide star is marked with a cross.}
    \label{swanfield}
    \end{center}
\end{figure}    

SWAN is the AO-assisted payoff of these seeing-limited preliminaries.
Having already characterized large samples of objects in our bright star 
fields, we targeted them with NACO on the VLT in order to exploit the present 
generation of AO technology for galaxy evolution studies.  NACO comprises the 
NAOS Shack-Hartmann AO module (Rousset et al. \cite{rousset}) mated with the 
CONICA near-infrared camera (Lenzen et al. \cite{lenzen}).  Our choice of NACO
observing mode was dictated by our desire to differentiate SWAN from previous 
HST/NICMOS surveys.  First, we chose to prioritize survey area over depth, 
thereby improving SWAN's sensitivity to rare objects and its robustness against 
cosmic variance.  Second, we chose to image in $K_\mathrm{s}$, where NICMOS is 
less sensitive than in $J$ and $H$, thus making SWAN preferentially sensitive 
to red objects.  Use of NACO's $0.054\arcsec$ pixel scale (to maximize field 
of view) and the Strehl ratios of 30--60\% typically achieved in 
$K_\mathrm{s}$ thus result in images that are undersampled.  Each NACO 
pointing provides only a usable $\sim 0.75\,\mathrm{arcmin}^2$ of the full $55.5\arcsec 
\times 55.5\arcsec$ detector area, due to losses from dithering and the 
central star (see, e.g., Fig.~\ref{swanfield}).  Nevertheless, the anticipated 
survey area that will result from assembling 42 such images will be -- at 
$\sim 30\,\mathrm{arcmin}^2$ -- some six times larger than the NICMOS survey 
of the HDF and flanking fields in $J$ and $H$ (Dickinson \cite{dickinson99}, 
Dickinson et al. \cite{dickinson00}).  
In the ecology of near-IR surveys, SWAN aims to occupy a niche combining the 
high angular resolution of a space-based survey with the shallower depth and 
wider area of a ground-based survey, thereby probing sources that are compact, 
faint, red, and rare more effectively than any other survey to date.
First results from an initial analysis of nine SWAN AO fields are presented 
in Baker et al. (\cite{baker05}), where NACO imaging is seen to 
detect a population of compact galaxies that cannot be identified as such in
seeing-limited data. 

To extract full information about the morphology of the sources detected in 
the SWAN images by taking into account the effects introduced by the PSF, it 
is necessary to develop a reliable model of the off-axis PSF in each of the 
fields. This task is not easy, however, as the AO PSF changes quickly in both 
time and position on the frame; in our case it is made even more difficult by 
the distinctive attributes of the SWAN observing strategy.  First, 
because of their high Galactic latitudes, relatively few point sources 
are present in each of the fields, so that very few objects can be used as 
references to constrain how the PSF varies off-axis.  In addition, in order to 
be background-limited so that the aim of the survey to detect faint galaxies 
can be realized, the on-axis star is always saturated on the science frames, 
meaning the on-axis PSF has to be obtained with separate unsaturated 
exposures.  Finally, NACO has only very rarely been used to image 
stellar fields with the (undersampled) $0.054\arcsec$ pixel scale and the 
visible wavefront sensor; as a result, the VLT archive contains no suitable  
stellar fields that can help characterize the properties of the PSF as far 
off-axis as the SWAN images extend. 

Given the above considerations, our data obviously pose several 
challenges for the recovery of intrinsic source parameters. Indeed,  
the same challenges will be faced by {\it any} extragalactic AO 
survey that relies on NGS or LGS wavefront sensing, and that 
efficiently builds up statistical samples of faint field galaxies by including
a wide area around each bright star position. It is thus of general interest
to devise a strategy to account for anisoplanaticism in deep, wide-field AO 
imaging.

\section{Other proposed methods to quantify anisoplanaticism} \label{methods}

There have been several methods proposed in the literature for
estimating or calculating the off-axis PSF in an AO field.
In this section we summarize these methods, and explain why they
are not ideal for deep wide-field AO imaging with current instrumentation.

A theoretical analytical expression to model PSF variation and
anisoplanaticism  
has been derived by Fusco et al. (\cite{fusco}), who validated their
method on both simulated and experimental data, leading to a reduction
in the error on the magnitude estimation in stellar fields from more
than 30\% to only 1\%.
They showed that the total optical transfer function (OTF) is simply the
product of the on-axis OTF with an anisoplanatic OTF.
The on-axis OTF can be readily obtained, for example from real-time data
accumulated by the AO system or a measurement of the on-axis PSF.
The anisoplanatic OTF, however, requires an independent measurement of
how the turbulence in the atmosphere depends on altitude, as
quantified by the atmospheric refractive index structure constant
$C^2_n$.
In their example case this was obtained by balloon probes, but similar
measurements are not usually available. 

In a similar vein, the possibility of using AO measurements together
with simultaneous scintillation detection and ranging 
(SCIDAR) measurements to
reconstruct the $C^2_n$ profile and hence the off-axis PSF, was
investigated by Wei\ss\,et al. (\cite{weiss02a,weiss02b}).
While such measurements were made and analyzed, the idea was never
brought to fruition and no such system has been installed as a
permanent facility instrument.

An alternative semi-empirical approach was proposed by 
Steinbring et al. (\cite{steinbring}), based on calibration images of
dense stellar fields to determine the change in PSF with field position. 
Their results showed that this simple method reduces the error 
in the prediction of the FWHM of the PSF at large distances off-axis
from 60\% to only $\sim 20\%$.
However, obtaining a suitable calibration field is not an easy task
as atmospheric conditions can change rapidly.
Steinbring et al. were able to complete the observations necessary to
construct their mosaic images in less than 10\,minutes.
However, this approach is not practicable for even moderately 
deep-field observations. Our SWAN observations, for example, typically 
have integration time (without overheads) of 60\,minutes.
The additional time needed to switch continually between science and
calibration fields would be prohibitive -- even assuming that a suitable
stellar field with a guide star of comparable brightness can be found not
too far from the science field and also at a similar airmass.
Furthermore, the authors point out that variations of up to 50\% in
the measured on-axis Strehl ratio ultimately limit the accuracy of the
results.

A third approach to the problem was investigated by 
Tristram \& Prieto (\cite{tristram}).
They parameterized the PSF using an elliptical Gaussian and a decaying
elliptical exponential function.
Starting with 11 parameters, they were able to eliminate all but four,
which would need to be derived from stars in the science field. 
At high Galactic latitudes, the difficulty is that the number of point 
sources will be very limited in any science fields. To constrain 4 
parameters reliably, the authors used 20 point sources, while our SWAN  
images always have fewer point sources per field 
-- in a number of cases there are fewer than four -- that are 
usually too faint for a detailed PSF fitting as is required here.

Given that it is not practical to apply any of these methods to SWAN
or similar data, in the next Section we propose an alternative method 
suited to cases where there are very few unresolved sources in the 
science field (and those which do exist are often rather faint), and where
one does not necessarily need a highly precise estimate of the PSF.

\section{The PSF model for SWAN} \label{our}

To tackle the issue of anisoplanaticism in the SWAN data and to model 
an accurate PSF for every galaxy we detect we have developed 
a parametric model of the variations of the PSF across the field of view. 
This kind of approach is particularly convenient, as it uses only
easily available data and makes no uncertain assumptions about the
stability of the isoplanatic angle during any given night. 
Although the PSF reconstruction is not perfect, we will
show that it is accurate enough for our purpose of measuring galaxy
morphologies.

According to a theoretical analysis of the anisoplanatic effect in AO
systems (e.g., Fusco et al. \cite{fusco}; Voitsekhovich \& Bara
\cite{v&b}), the off-axis PSF can be expressed as a convolution
between the on-axis PSF of the guide star with a spatially variable
kernel, i.e.,
\begin{equation}
        \textrm{PSF}(\textbf{\textrm{f}},\theta) = 
                \textrm{PSF}(\textbf{\textrm{f}},0) \otimes 
                        \textrm{K}(\textbf{\textrm{f}},\theta)
\end{equation}
\begin{table} 
\label{calibtab}
        \begin{center}
        \begin{tabular}{l c c c c}
        \hline 
        \hline \noalign{\smallskip}
        Field & Date & Filter &  Exp. Time & Strehl \\
         &  &        &  (min) & (\%) \smallskip \\
        \hline \noalign{\smallskip} 
        NGC 6809 & 15.12.02 & $K_\mathrm{s}$ & 60 & 28 \\
        NGC 6752 & 06.09.03 & $K_\mathrm{s}$ & 12 & 27 \\
        \hline
        \hline
        \end{tabular}
        \end{center}
        \caption{Observations of PSF calibration fields. 
        The Strehl ratio is that measured on-axis in the short exposure
        frames taken before and after the deep science exposures.}
\end{table}
We will make the extreme assumption that the model of the kernel 
$\textrm{K}(\textbf{\textrm{f}},\alpha)$ is sufficiently well
constrained that the difference in the kernel from one science field to
the next is reduced to a single parameter, namely the isoplanatic
angle. This is the most convenient solution for fields with 
sparse point source populations and, as we show, is able to reproduce a 
reasonable approximation to the PSF at any given location in 
the field of view using just the the on-axis PSF and a few reference point 
sources in the science field as calibration.

For each SWAN field the on-axis PSF was monitored by taking short
unsaturated images of the bright star both before and after (and in
some cases half-way through) the deep science exposure.
These short exposure frames were taken in the $K_\mathrm{s}$ band, 
IB\_2.21 ($\lambda_c = 2.210\,\mathrm{\mu m}$, FWHM = 0.060\,$\mu$m) or
NB\_2.17 ($\lambda_c = 2.166\,\mathrm{\mu m}$, FWHM = 0.023\,$\mu$m) filters.
The integration time varied in the range 0.5--2.0\,sec depending on
the filter used.
In addition, observations of galactic star cluster fields were planned 
periodically in order to build up a database of anisoplanatic PSFs. 
At the time of writing only two such calibration fields had been
observed (see Table \ref{calibtab}), but between them and the 7 SWAN fields 
with the highest numbers of point sources (see end of this section), 
included in Fig. 2, they still provide nearly 80 different point sources. 
These fields will be used in the following analysis as a test bench to
derive a suitable parametric model for the off-axis PSF.

The images were reduced using PC-IRAF version 2.11.3.
The presence of the bright star in the center of a field less than
$1'$ across made the data reduction a little more complex then usual,
requiring two iterations of object masking and sky subtraction to
avoid over-subtraction from very extended faint scattering 
around bright objects. 
For further details about data reduction see Cresci et al. 
(\cite{cresci}).

\begin{figure}
    \begin{center}
     \resizebox{\hsize}{!}{\includegraphics{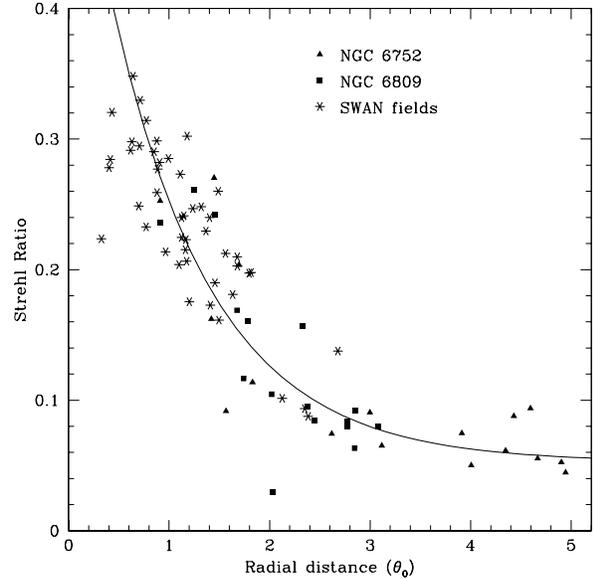}}   
    \caption{Isoplanatic angle scaling. The Strehl ratios of point 
    sources in the two calibration fields and 7 SWAN fields 
    (SBSF\,14, SBSF\,15, SBSF\,18, SBSF\,24, SBSF\,27, SBSF\,28, SBSF\,41) 
    are plotted against their 
    radial distance $\alpha$ from the AO guide star (in 
    units of the corresponding isoplanatic angle). 
    The black solid line is the theoretical curve used for fitting.}
    \label{sr}
    \end{center}
\end{figure}    

The isoplanatic angle of these fields was measured 
by considering -- independently for each field --
the Strehl ratio of the detected stars as a function of radius from
the guide star.
To these data we fit the theoretically expected function (e.g., Beckers
\cite{beckers}; Roddier \cite{roddier}):
\begin{equation}
\label{isoplan}
   \mathrm{SR} \propto 
   \textrm{exp}\left[-\left(\frac{\Delta\theta}{\theta_0}\right)^{\frac{5}{3}}\right]
\end{equation}
where $\Delta\theta$ is the offset angle from the guide star on the sky and
$\theta_0$ the isoplanatic angle, defined as the separation angle at which the
Strehl ratio has degraded by a factor $e$ with respect to the on-axis
value. 
Although in a strict sense this refers to a telescope with infinite
aperture (e.g., Beckers \cite{beckers}), for our implementation this
definition is adequate.
A constant term was added to Eq.~\ref{isoplan} in order to match 
the observed data.
The resulting best fitting values for $\theta_0$ are 9.30\arcsec\ and
6.61\arcsec\ for NGC\,6809 and NGC\,6752 respectively.
These are surprisingly small given the mean isoplanatic angle of 
the other seven SWAN fields of $\sim$17\arcsec.
However, they do provide a larger range in $\theta_0$  over which our
model can be constrained.
The radial distances of all the point sources in the two calibration fields 
and in the 7 observed SWAN fields could then be normalized to their 
isoplanatic angles. 
This is shown in Fig. \ref{sr}, where the Strehl ratios of the point
sources in these fields are plotted against their rescaled separations from 
the AO guide star.
It can be seen that all the points now lie on the same theoretical
distribution, showing that the isoplanatic angle is the source of the
principal variation between the different fields.

\begin{figure*}
        \begin{center}
        \resizebox{0.7\textwidth}{!}{\includegraphics{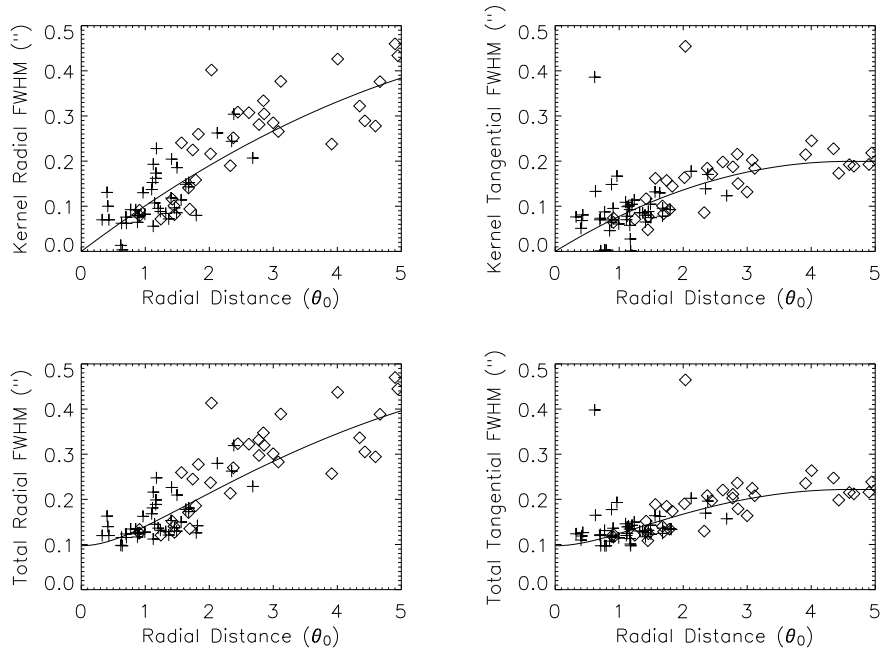}}   
        \caption{\textit{Upper panels}: FWHM fits to the kernel. The radial 
        (left panel) and tangential (right panel) FWHMs of the kernel to 
        be convolved with the on-axis PSF to reproduce the observed PSF for 
        the point sources in the calibration fields (diamonds)
        and in the seven SWAN fields of Fig.~\ref{sr} (pluses), 
        along with the best 
        fitting second order polynomial. \textit{Lower panels}: 
        The total FWHMs of the observed point sources (radial at left,  
        tangential at right) as a function of radial distance 
        $\alpha$ from the guide star in units of the corresponding 
        isoplanatic angle $\theta_0$, compared to the best fitting model 
	for the kernel convolved with the guide star's FWHM.} 
    \label{fwhm}
    \end{center}
\end{figure*}

Our model is motivated by the fact that the wavefront error in the
anisoplanatic kernel is dominated by the tip-tilt terms.
This is borne out in practice, since the major degradation of the PSF
observed in the calibration field images is an increasing radial elongation
towards the guide star as one moves further off-axis.
This suggests that, as a first approximation, one can represent the 
anisoplanatic kernel $\textrm{K}(\textbf{\textrm{f}},\alpha)$ with an   
elliptical Gaussian elongated towards the guide star.
For each of the point source in both of the calibration
fields, we therefore derived the best fitting radial and tangential FWHMs
of the elliptical Gaussian kernel to be convolved with the  
on-axis PSF in order to reproduce the observed off-axis PSF (see
Fig. \ref{fwhm}).  \\
The fit was obtained by minimizing the sum of the squared
difference between the observed star and models with different kernel
FWHMs, weighted by the flux in each pixel of the star frame (in order 
to optimize the fit in the PSF core) and including all pixels brighter 
than 1\% of the star peak. As the noise is dominated by the background, which 
is constant over each star, a noise term was not included in the expression, 
because it would have added just a constant scaling factor.

\begin{figure}
    \begin{center}
     \resizebox{0.4\textwidth}{!}{\includegraphics{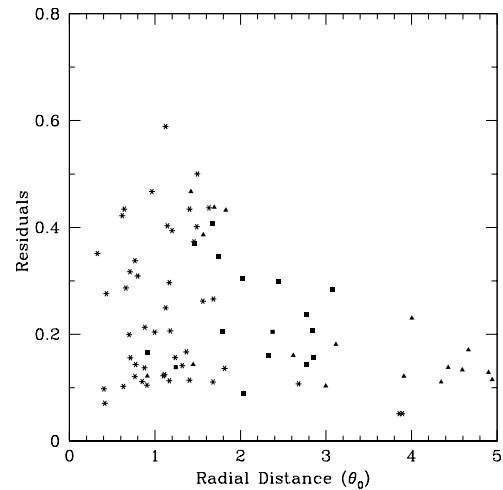}}   
    \caption{Residuals from the PSF fitting with the elliptical kernel for the 
    point sources in the two calibration fields, NGC\,6752 (triangles) 
    and NGC\,6809 (squares), and in the 7 SWAN fields of Fig.~\ref{sr} (stars). The 
    residuals plotted are defined in Eq.~\ref{reseq}.}
    \label{residuals}
    \end{center}
\end{figure}

The residuals of the fitting are shown in Fig. \ref{residuals} as a
function of separation from the guide star. 
The residuals are defined as follows:
\begin{equation} \label{reseq}
       \textrm{Residuals}=\frac{\sqrt{\sum_i((\textrm{Star}_i-\textrm{Model}_i)^2 \cdot 
                \textrm{Star}_i)}}{\left(\sum_i(\textrm{Star}_i)\right)^{1.5}} \cdot 100
\end{equation}
where \textit{Star} is extracted from the NACO data and \textit{Model}
is the best fitting kernel convolved with the on-axis PSF, so that the
numerator is the quantity minimized during the fit, which was then
normalized using the star brightness. 
It is clear that not all of the stars are equally well represented by
the model, as expected due to the very simple kernel adopted to
reproduce the PSF. 
However, the residuals do not depend strongly on the field in which the star
lies (see Fig.~\ref{residuals}) suggesting that the approach is robust 
to different conditions. The contribution of the noise 
to the residuals could be significant in the fainter point sources 
in the SWAN fields.

The FWHMs obtained along the radial and tangential directions were
then fitted empirically as functions of the radial distance from the
AO guide star. 
The function chosen was a second order polynomial, which was forced to 
a constant after reaching a maximum. 
Using higher order polynomials (or other analytical functions) just
increased the number of free parameters without improving the fit. 
The final best fitting functions obtained are

for the radial axis:
\begin{equation}
        \textrm{FWHM}_r(\arcsec)=\left\{ \begin{array}{ll}    
                0.107 \cdot \alpha - 0.00594 \cdot \alpha^2 & \textrm{if}\ \alpha \leq 9.1\\
                0.481 & \textrm{if}\ \alpha > 9.1
        \end{array} \right.
\end{equation}
and for the tangential axis:
\begin{equation}
        \textrm{FWHM}_t(\arcsec)=\left\{ \begin{array}{ll} 
                0.0871 \cdot \alpha - 0.00951 \cdot \alpha^2 & \textrm{if}\ \alpha\leq 4.6\\
                0.199 & \textrm{if}\ \alpha > 4.6
        \end{array} \right.
\end{equation}
where $\alpha=\theta/\theta_0$ is the radial distance from the guide star 
rescaled using the isoplanatic angle. 
We notice that, although the method could be in principle applied to every AO
system, the derived parameters were derived specifically for our SWAN dataset, 
and they have to be re-calibrated in order to use the same PSF model for any 
different adaptive optics system, instrument setting and operating conditions.\\
The variation of the total FWHM -- i.e., the kernel convolved with the on-axis 
PSF -- along both radial and tangential directions for the point sources 
in the calibration field is shown in the lower panels of Fig.~\ref{fwhm}. 
It can be seen how the total FWHM has a behavior consistent with 
what is observed for other AO PSFs (e.g., Flicker \& Rigaut \cite{flicker}). 

Using the derived parametric model for the kernel, it is now possible
to build the model PSF for any position in any of the
SWAN fields with only an estimate of the isoplanatic angle in the
respective field.
The isoplanatic angle is derived by fitting the Strehl ratio of the
few point sources detected in each science frame as a function of
radius with Eq.~\ref{isoplan}. 
We have used SExtractor (Bertin \& Arnouts \cite{bertin}) 
to identify the sources in our SWAN NACO images, and  
as reference point sources for fitting the Strehl ratio, we use those
with SExtractor 
stellarity index $\mathrm{SSI} \geq 0.9$ (see Fig.\ref{swanfield}). 
The derived isoplanatic angle is then used to rescale 
the radial distance from the guide star and to build the model PSF.

\begin{figure}
    \begin{center}
     \resizebox{\hsize}{!}{\includegraphics{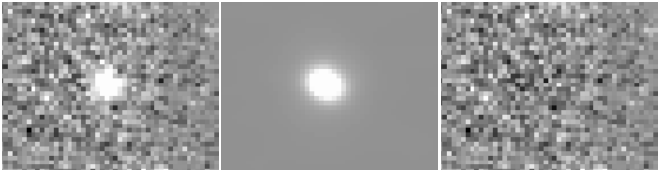}}   
    \caption{Comparison between the true and model PSFs for a point
    source in NGC~6752.
    The left panel shows the original star, the central panel the
    model for the  
    corresponding position on the detector, and the right panel the
    residuals after subtracting the two. The star is located at
    26.8\arcsec\ from the AO guide star.} 
    \label{fitpsf}
    \end{center}
\end{figure}
We additionally compared these results to those that could be obtained 
using PSFs generated by the PAOLA software 
(\textsf{http://cfao.ucolick.org/software/paola.php}), with an
appropriate isoplanatic angle.
PAOLA (Performance of Adaptive Optics for Large Apertures) is a set of 
functions and procedures written in IDL for calculating the performance 
of an AO system installed at the focus of an astronomical telescope. 
It relies on an analytic expression of the power spectrum of the corrected phase 
and its relation to the AO OTF.  It therefore assumes that the AO system 
is ``perfect'', failing to reproduce second order effects due to 
imperfections of the system. A set of PSFs were generated at different 
distances from the guide star, using in the model average parameters
for the VLT and NACO. The radial distances of the PSFs were then
rescaled to match the isoplanatic angle observed in the different
fields. 
However, the match with the true PSFs was no better -- and often worse
-- than that obtained with our parametric model, 
in particular worsening for increasing distances from the guide star.

\begin{figure*}
        \centering
        \begin{minipage}[c]{1.0\textwidth}
        \centerline{\hbox{
                  \psfig{file=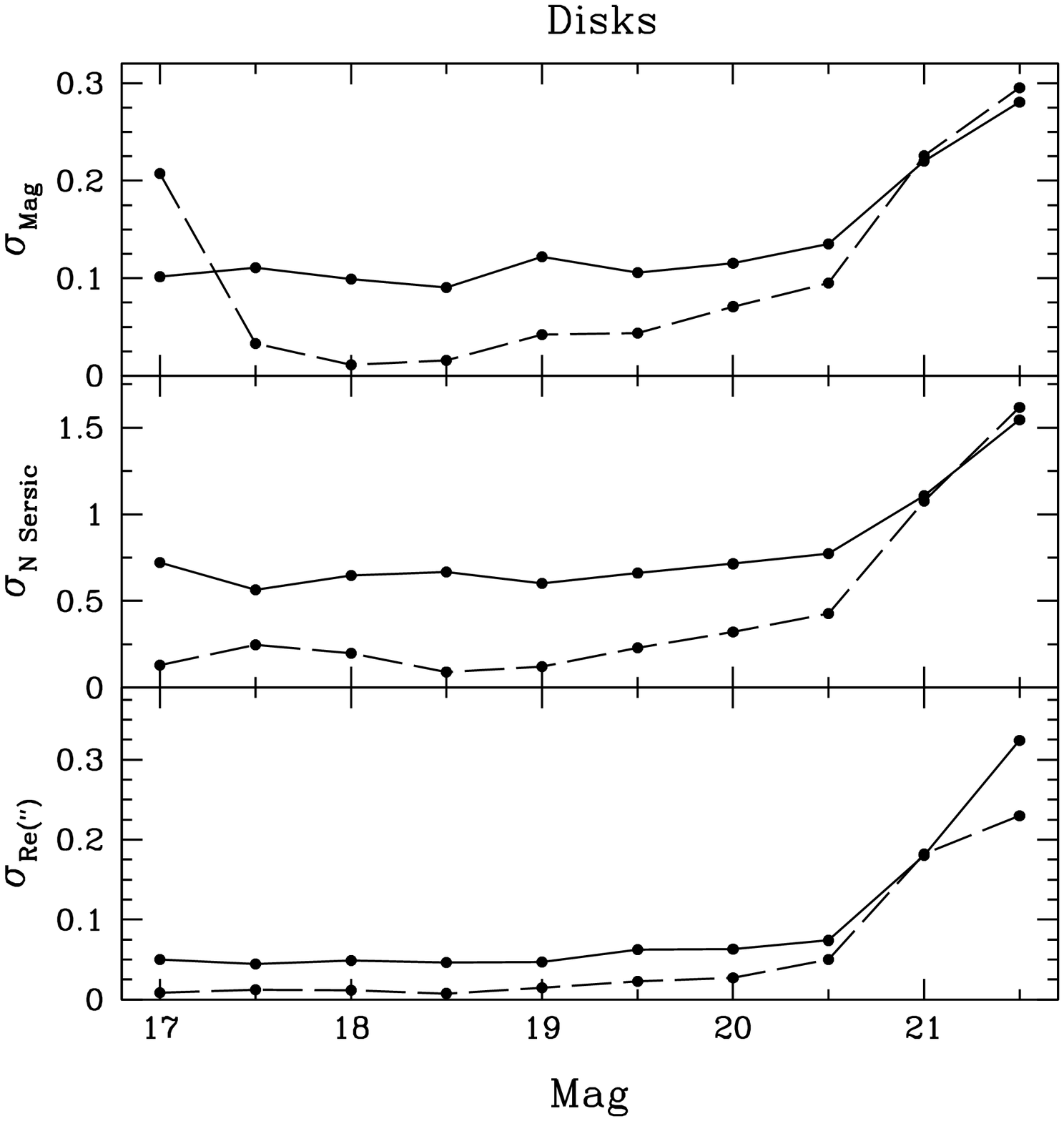,width=9.0cm}
                 \hspace{0.0cm}
                  \psfig{file=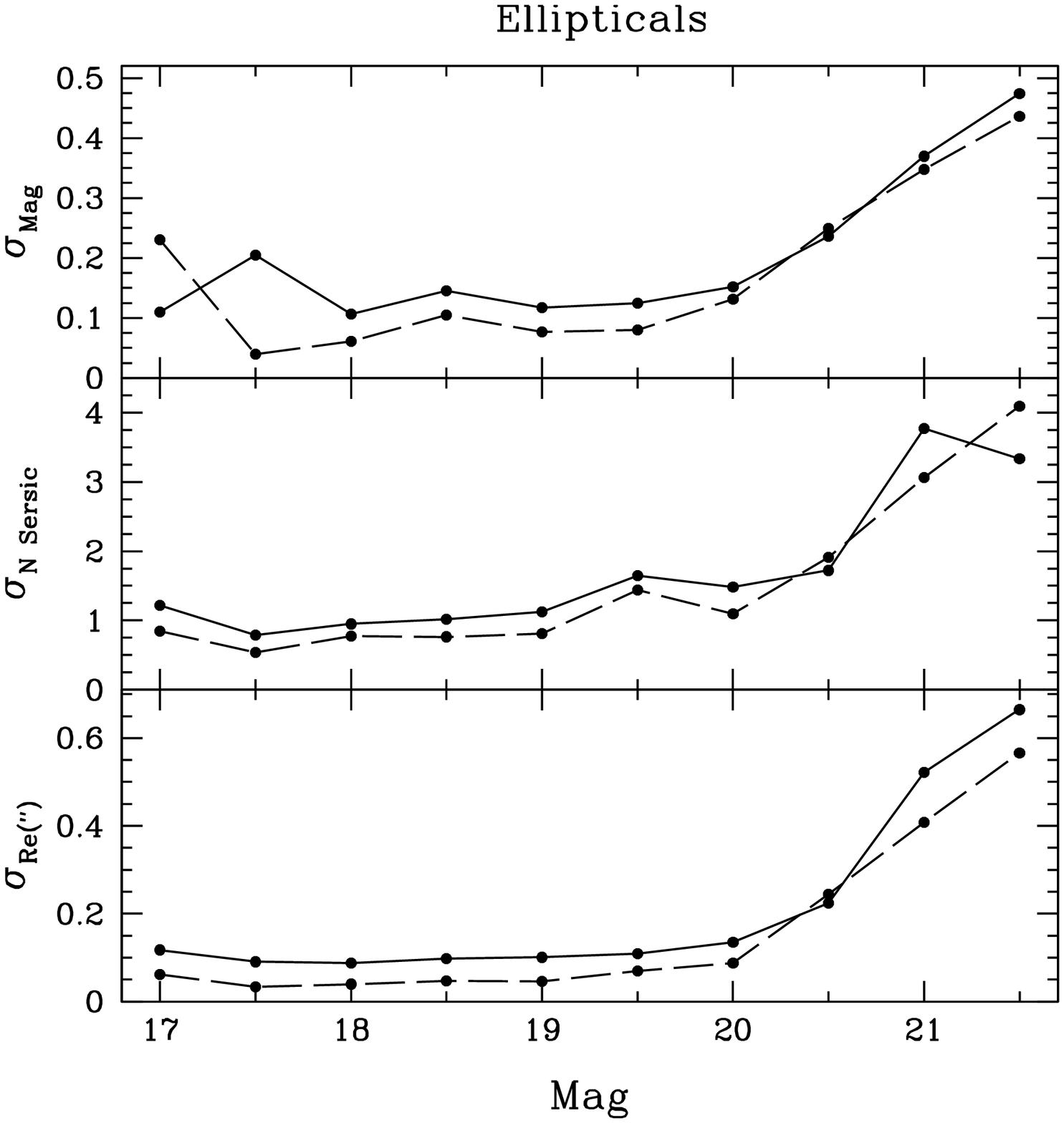,width=9.0cm}
        }}
        \caption{The \textit{left panels} show the RMS of the magnitude, 
        S\'ersic index and effective radius obtained from the GALFIT fits of the 
        simulated profiles with S\'ersic index $n = 1$ as a function 
	of the input magnitude. 
        For each point, 100 galaxies with 10 different PSF were
        used (see text for details).
        The solid lines are the RMS obtained using the model PSF as an input for 
        GALFIT, while the dashed ones are obtained with the original NACO PSF. 
        The \textit{right panels} show the results obtained for simulated 
        profiles with S\'ersic index $n = 4$.
        In both cases it is clear that the parameters can be reliably
        recovered up to $K_\mathrm{s} \sim 20$.}  
        \label{sigmasimul}
        \end{minipage}
\end{figure*}

\section{Testing the PSF model with simulations of galaxy profiles} 
\label{simul}

The PSF model developed in the previous section was carefully tested
to see how well it performs in terms of recovering the correct morphological 
parameters of simulated galaxy profiles.
The galaxy models were built by convolving a S\'ersic (\cite{sersic}) profile,  
\begin{equation}
        I(R) = I(R_e) \times e^{\left( -b_n \times [(R/R_e)^{1/n} - 1]\right)}
\end{equation}
(where $R_e$ is the effective radius, $n$ is the S\'ersic index and $b_n$ 
is a constant that varies with $n$)  
with a true PSF extracted from either a calibration or a SWAN field.
Our tests are particularly robust since we can compare the
parameters recovered using the model PSF not only with those used as
input for the simulations, but also with those recovered using the
true PSF used to convolve the simulated profile.
The primary aims of the simulations were to determine how well it was
possible to estimate as a function of magnitude
(1) the S\'ersic index $n$ -- and in particular whether it was possible
to separate disk and elliptical profiles, and
(2) the effective radius $R_e$ of the profile, which indicates the
size scale of the galaxy.

We used S\'ersic index $n = 1$ for disk-like galaxies and 
$n = 4$ for elliptical-like galaxies. 
For both types we ran several sets of simulations, each including 100
galaxies, at fixed magnitudes ranging from 17--22\,mag in the $K_\mathrm{s}$ 
band, with the same pixel scale used in the SWAN images. 
The noise was set to the level expected for NACO integration time 
of 60\,min for all the galaxies, since this is the typical integration 
time for a SWAN field. The inclination and position angle  
were random for each simulated profile, while the distribution of
effective radii $R_e$ was chosen to roughly reproduce that observed in
the real data -- i.e. 40\% of the objects having $R_e = 0.1\arcsec$, 
20\% each with $R_e = 0.2\arcsec$ and $R_e = 0.3\arcsec$, 
and 10\% each with $R_e = 0.4\arcsec$ and $R_e = 0.5\arcsec$. 
Ten different PSFs extracted from point sources in the calibration or
SWAN fields were used for each set of 100 galaxies.

To recover the galaxy morphological parameters, we used GALFIT (Peng
et al. \cite{peng}), a widely used software package which fits an image of a
galaxy and/or point source with one or more analytic functions.
For our simulations we used GALFIT to fit a single S\'ersic profile,
leaving as free parameters the center of the galaxy, the S\'ersic index
$n$, the effective radius, the magnitude, the position angle and the
axis ratio.
The software needs an image of the PSF as input in order to fit the
galaxy profile, both to deconvolve the original galaxy image and to
convolve the derived model.
For each simulated galaxy we ran GALFIT using both the true PSF from
the NACO data -- i.e., that which had been used to create the profile
-- and the corresponding model PSF, in order to compare the results.

The results of the simulations are shown in Fig.~\ref{sigmasimul},
where the uncertainties in the derived magnitude, S\'ersic index, and
effective radius are plotted as a function of magnitude for both
disk-like and elliptical-like profiles.
The results obtained using both the model PSF (solid lines) and
true PSF (dashed lines) as input to GALFIT are shown, and it can be
seen that these are in general very similar.
Since using the true PSF is the best that can be done, we conclude
that -- in the application here of deriving the morphological parameters 
of galaxies -- our model is a reasonable and valid approximation to
the true PSF.

\begin{figure}
    \begin{center}
     \resizebox{0.45\textwidth}{!}{\includegraphics{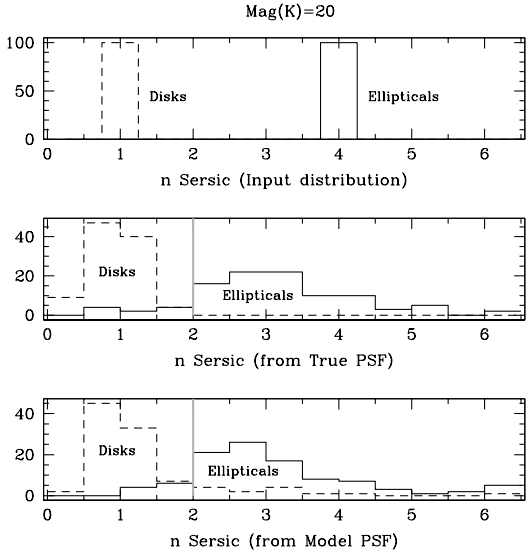}}   
    \caption{Disk/elliptical separation using GALFIT measurement of S\'ersic 
    index for simulated profiles with $K_\mathrm{s} = 20$. The 
    \textit{upper panel} shows the input distributions of S\'ersic index, 
    while the \textit{central panel} 
    shows the results using the true NACO PSF used to build the profile 
    as an input for GALFIT, and 
    the \textit{lower panel} what is obtained with the model PSF. 
    The discriminating value for the S\'ersic index $n = 2$ (see text) is 
    shown as a gray line.}
    \label{diskbulge}
    \end{center}
\end{figure}

\begin{figure}
    \begin{center}
     \resizebox{0.45\textwidth}{!}{\includegraphics{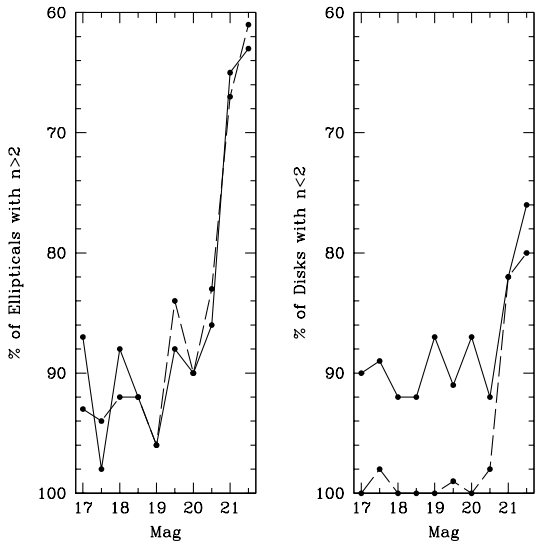}}   
     \caption{The fraction of ellipticals recovered with $n>2$ 
     (\textit{left panel}) and the fraction of disks recovered with a 
     S\'ersic index $n < 2$ (\textit{right panel}) as a function of the 
     input magnitude. The dashed lines show the results obtained using the 
     true NACO PSF, while the solid lines the ones obtained with the 
     model PSF.}
    \label{wrongn}
    \end{center}
\end{figure}

The most significant discrepancy between the two sets of results is a
larger RMS in the S\'ersic index for the simulated disk galaxies when
the model PSF was used in GALFIT.
In addition, the RMS for the S\'ersic index is quite large ($\sim 1$)
for elliptical galaxies, but this time for both the model and true PSFs. 
In principle this rather large uncertainty on the S\'ersic index could
restrict our ability to separate the two populations of disk-like and
elliptical-like galaxies.
However, in practice, the large variance for the ellipticals is due to
a long tail to high values of $n$, as apparent in Fig.~\ref{diskbulge}.
We have found that starting with two populations of simulated
galaxies with S\'ersic index $n = 1$ and $n = 4$ respectively, only a small
fraction (that increases with the magnitude of the galaxies, see 
Fig.~\ref{wrongn}) of disks was recovered with $n > 2$, and few ellipticals 
were recovered with $n < 2$. 
The two classes can be clearly distinguished up to $K_\mathrm{s} = 20.5$, where
90\% of the disks are recovered with S\'ersic index $n < 2$ and 90\% of the 
ellipticals are recovered with S\'ersic index $n > 2$, using  the model PSF 
in GALFIT (see Fig.~\ref{diskbulge}).
As a result, it is possible to set a threshold of $n = 2$ that is able to
discriminate between the two populations with a high degree of success.
The fraction of disks recovered with a S\'ersic index $n < 2$ and the 
fraction of ellipticals recovered with $n > 2$ are shown in Fig.~\ref{wrongn} 
as a function of the magnitude, using both the model and true PSF. 
As the correct fitting of the S\'ersic index is limited by the signal to noise 
of the sources, in order to correctly discriminate between disks and 
ellipticals at a fainter magnitude $K_\mathrm{s}$, the integration time should  
be scaled as $\Delta t\,(\mathrm{hr})=10^{[0.8\cdot(K_\mathrm{s}-20.5)]}$.  

As an additional check of the usefulness of the model
PSF, we repeated the experiment described above but giving GALFIT the
observed on-axis PSF instead of either the model or true PSF.
The outcome was that, even at relatively bright magnitudes $K_\mathrm{s} 
\simeq 17$, all the galaxies with a S\'ersic index $n = 4$ were actually 
recovered with an index of $n < 2$, making it impossible to discriminate 
between the two populations.
This emphasizes the important result of our simulations that it is only
possible to discriminate between disk galaxies and elliptical
galaxies using an anisoplanatic PSF, 
and that using our very simple PSF model is nearly as good as using
the true PSF.

\section{First results from five deep science fields} \label{5results}

Prompted by the results on the simulated galaxy profiles, we have
begun using our PSF model coupled with GALFIT to derive morphological 
parameters of the galaxies detected in a few of the SWAN survey fields 
in order to gain some insight into what we should expect,
although clearly a full analysis requires 
the results of the dedicated redshift survey currently underway.
In this paper we present the first results obtained from the first
five fields, SBSF\,14, SBSF\,15, SBSF\,18, SBSF\,24 and SBSF\,41, 
all of them already included in the preliminary discussion in Baker et al. 
(\cite{baker05}).
The observations are summarized in Table~\ref{obstab}, and the data
were reduced using the same procedure adopted for the calibration 
fields (see Sect. \ref{our}).

\begin{table} 
\label{obstab}
        \begin{center}
        \begin{tabular}{l c c c c c}
        \hline
        \hline \noalign{\smallskip}
        Field & Date & Filter &  Exp. Time & CE & $\theta_0$ \\
              &      &        &  (min)     & (\%)   & (\arcsec)
        \smallskip \\
        \hline \noalign{\smallskip}
        SBSF\,14 & 15.12.02 & $K_\mathrm{s}$ & 60 & 62 & 11.9\\
        SBSF\,15 & 17.12.02 & $K_\mathrm{s}$ & 60 & 66 & 21.8\\
        SBSF\,18 & 17.12.02 & $K_\mathrm{s}$ & 40 & 68 & 21.2\\
        SBSF\,24 & 21.03.03 & $K_\mathrm{s}$ & 60 & 36 & 11.8\\
        SBSF\,41 & 14.06.03 & $K_\mathrm{s}$ & 60 & 53 & 19.7\\
        \hline
        \hline
        \end{tabular}
        \end{center}
        \caption{Observations of the five SWAN fields included 
        in the analysis here. \textit{CE} is the encircled energy 
        reported by the adaptive optics system in real time. 
        The isoplanatic angle $\theta_0$ 
        is that derived by the fit using point sources in the 
        fields (see Sect.~\ref{our}).}
\end{table}

The performance obtained in these first SWAN fields compares
favorably with that obtained by previous AO observations of faint field 
galaxies, e.g., Larkin et al. (\cite{larkin00}) and 
Glassman, Larkin, \& Lafreni\`ere (\cite{glassman}). They obtained 
12 $H$-band ($1.6\,\mathrm{\mu m}$) images of disk galaxies at 
$z \sim 0.5$ using the AO system on the Keck II telescope. 
They obtain FWHMs ranging from 0.050\arcsec\ 
to 0.15\arcsec, and Strehl ratios of 1\%--20\%. Due to the very small 
field of view (4.5\arcsec) only one galaxy could be observed for each 
frame. The PSF was reconstructed using dedicated observations of 
faint off-axis stars with comparable offset, but different position 
angle, from the guide star. 
Due to the very low S/N obtained, the extraction of morphological 
parameters was run under several assumptions to limit the number 
of free parameters in the fit of a bulge and a disk to each galaxy. 
In addition, the AO galaxy profiles only had reasonable S/N out to 
1\arcsec--2\arcsec\ radius, thus limiting the constraints on the outer 
portions of the disks.
Steinbring et al. (\cite{steinbring04}) imaged with the same AO system 
three galaxies selected because they had been observed previously 
with {\em Hubble Space Telescope} at optical wavelengths. However they 
conclude that higher S/N was needed in the AO images in order 
to match the HST data, and the results were limited 
due to poor constraints on the AO PSF.

\begin{figure*}
        \begin{center}
        \resizebox{0.8\textwidth}{!}{\includegraphics{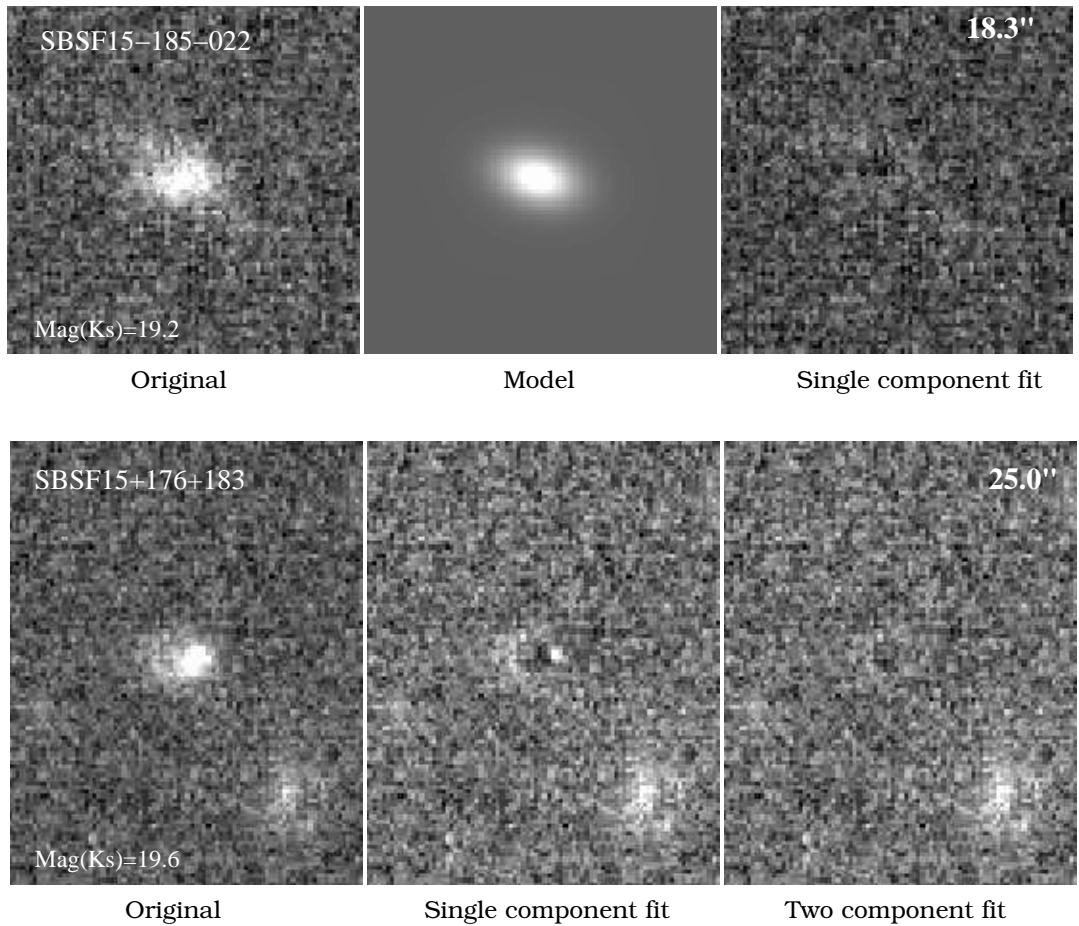}}   
        \caption{Examples of single and multiple components fits by GALFIT of 
        SWAN galaxies using the model PSF. In all images North is up and East 
        is left. The distance of the galaxy from the guide star is shown in the 
        upper right corner of each panel. For both galaxies, the $K_\mathrm{s}$ 
        magnitude and the source identifier according to Baker et al. 
        (\cite{baker}) are reported.  
        \textit{Upper panel}: single component 
        S\'ersic fit ($K_\mathrm{s} =19.2$; 
        $R_e = 0.4\arcsec$; $n = 0.6$; $b/a = 0.6$; PA = 75). The left image is 
        the original galaxy, the GALFIT best fitting profile is in the middle, 
        and the residuals of the subtraction are on the right. 
        \textit{Lower panel}: comparison of the 
        residuals between a single 
        component fit (in the middle) and a two component fit (on the right), 
        while the original galaxy is shown on the left. The final best 
        fit was obtained with a disk component ($K_\mathrm{s} = 19.6$; $R_e = 
        0.4\arcsec$; $n = 1.0$; $b/a = 0.7$; PA = 76) and a point source 
        ($K_\mathrm{s} = 21.4$) in the center.}
    \label{galfit}
    \end{center}
\end{figure*}

\subsection{Source detection and modeling}

In each SWAN field, sources were detected using SExtractor, with the
appropriate parameters set to provide a positive detection for objects
brighter than $1.5\,\sigma$ per pixel over an area of more then
3~pixels. 
To improve the detection of faint sources we used a Gaussian 
filter ($\sigma=1.5$~pixels) to smooth the image. 
False detections at the noisy borders 
of the mosaic and on the spikes and the ghost of the bright guide star
were removed.
For the former, a mask which indicted the fraction of
the total integration time spent on each pixel was used; objects detected in
pixels below a specified threshold were rejected.
For the latter appropriate object masks were created.
The resulting coverage of these 5 fields was $3.2\,\mathrm{arcmin}^2$, within
which a total of 178 sources were detected down to a magnitude of
$K_\mathrm{s} \sim 23$ ($K_{\mathrm{AB}}\sim24.8$).
Of these, 146 were allocated a stellarity index $\mathrm{SSI} < 0.9$ by
SExtractor, and hence classified as galaxies.
The SExtractor classifications should be treated with caution
since they assume a constant PSF across each field. However, 
all the objects classified as stars by SExtractor lie on an upper 
envelope in a Strehl versus radial distance plot, i.e., they have the 
highest Strehl ratio among the sources at the same distance from 
the guide star,  
supporting their classification as point sources. 

We again used GALFIT to derive the morphological parameters of the
detected galaxies.
We used single component S\'ersic profiles, providing the output from
SExtractor as the initial guesses for magnitude, position, position
angle, and axis ratio.
For the deconvolution, a model PSF was created using the position of
the center according to SExtractor and scaled according to the isoplanatic
angle, which is shown for each field in Table~\ref{obstab}.
For each object, GALFIT was run twice, using as initial guesses for the
second iteration the outputs of the first iteration.

Some of the sources still showed bumps in the center or disk-like
structures in the residuals after the single component fit. 
These objects were re-fitted using either two S\'ersic components
(bulge~+~disk) or a S\'ersic component with an additional point source,
to improve the fit and minimize the residuals. 

A second iteration was also performed for close pairs or interacting
galaxies; we fit both of them simultaneously to avoid incorrect
background estimation or contamination from the companion. 
Some examples of single and multiple component fits are shown in 
Fig.~\ref{galfit}.
For galaxies fitted using multiple components, we report only the
values for the larger scale component;
the nuclear properties will be discussed elsewhere.

\subsection{Source morphologies}

\begin{figure}
\centering
        \resizebox{0.45\textwidth}{!}{\includegraphics{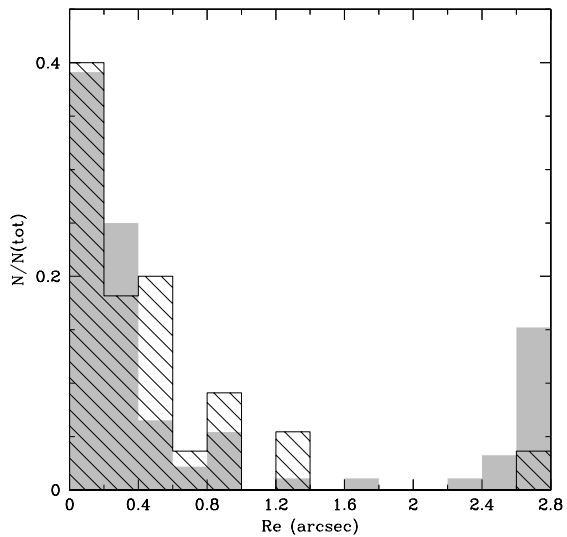}}
        \caption{Distribution of effective radii $R_e$ for the 55 sources
        detected down to $K_\mathrm{s} = 20$ in the five SWAN fields  
        observed with NACO (black hatched histograms), and for the 92 
	sources down to $K_\mathrm{s} = 20$ observed in in larger 
	fields centered on the same stars with SOFI (gray shaded histograms). 
	The distribution is not corrected for completeness.      
        For galaxies with more than one component, only the parameters
        of the larger scale (i.e., non-nuclear) one are used.} 
        \label{sumresults}
\end{figure}

The results obtained for the effective radii of
the 55 galaxies brighter than $K_\mathrm{s} = 20$ are shown in
Fig.~\ref{sumresults}.
While the redshifts of these objects are presently unknown, 
the magnitude-redshift relation of Cowie et al. (\cite{cowie})
 and the K20 survey (Cimatti et al. \cite{cimatti}) 
indicate that at $K = 20$ the median redshift is $z \sim 0.8-1$.
At this redshift, our spatial resolution of $0.1\arcsec$, which also
corresponds to the smallest effective radius bin, is equivalent to
only 500\,pc for typical cosmologies, hinting at the exciting
potential of this work.

Using $n = 2$ to discriminate between disks and ellipticals, we find there are 
24 elliptical-like and 31 disk-like galaxies, with uncertainties below
10\% on both numbers (see Section~\ref{simul}).
In fact, with similar numbers of each type found, the number of
ellipticals scattered into the disk-like category should roughly compensate
the number of disks scattered out of it, and vice-versa.

The distribution of effective radii is strongly peaked towards very
compact sources, even down to scales of 0.1\arcsec.
To check if this is due to a selection effect introduced by the AO
correction -- since adaptive optics data are inherently most sensitive
to compact sources -- we ran GALFIT on the same five fields, but using
the seeing 
limited observations made with SOFI (Baker et al. \cite{baker}). 
The same procedure was used, but a fixed PSF derived from the
unsaturated guide star in each SOFI science fields was provided to GALFIT
instead of the model PSFs.
Not all the galaxies detected in the NACO images are present in the
SOFI data, but on the other hand the SOFI spatial coverage is $\sim 5$
times larger, so that in the end we detected a comparable number of
objects: 181 in total, 92 brighter than $K_\mathrm{s} = 20$. The obtained 
distribution is shown in Fig.~\ref{sumresults} as a gray histogram. 
The distribution of $R_e$ shows similar behavior in both AO and seeing
limited data, suggesting that the observed distribution of $R_e$ in the SWAN 
fields is real and not due to selection effects. 
Nevertheless, the NACO distribution shows an excess of sources in the 
range $R_e = 0.4-1.2$ with respect to SOFI, indicating that the better 
resolution of NACO is able to resolve many sources that in the SOFI data 
appear merely ``compact'' (e.g., a bright elliptical galaxy that NACO correctly  
recovers with $n=4$ and $R_e=1\arcsec$ is recovered by SOFI for a seeing of 
$0.7\arcsec$ as a compact source with $R_e=0.1\arcsec$, while a disky galaxy 
with $n=1$ and the same effective radius is correctly recovered by both).

Finally, the axis ratio distribution shows the co-sinusoidal
distribution expected from random inclinations, except for a peak at 
$b/a = 0$ due to very compact galaxies for which GALFIT could not 
calculate the axis ratio, and hence which were arbitrarily assigned 
to this bin.

\subsection{Completeness}

We have derived a first estimate of the completeness
limit in our images using simulated galaxy profiles convolved with true 
NACO PSFs. 
This is a difficult task, as the completeness depends not only on the
brightness and morphology of the source, but even on the distance from
the guide star and the position on the frame. 
Therefore we derived a first estimate using only one reference PSF, a
star at $21.2\arcsec$ from the guide star in SBSF\,15. 
Fields with simulated galaxies of fixed magnitude and 
different $R_e$ (from $0.1\arcsec$ to $1.0\arcsec$) were created, and the sources
were recovered using the same SExtractor parameters as used for the
SWAN fields.
We find that at $K_\mathrm{s} = 20$ 
all the simulated elliptical-like galaxies were 
detected, while the disk-like galaxies were only 100\% complete for
effective radii smaller than $R_e = 0.6\arcsec$.
An average 50\% completeness limit -- i.e., the 50\% limit for 
galaxies with $R_e = 0.3\arcsec$ -- is $K_\mathrm{s} \sim 21.5$ 
($K_{\mathrm{AB}}\sim23.3$) for disk-like galaxies and $K_\mathrm{s} 
\sim 23$ ($K_{\mathrm{AB}}\sim24.8$) for elliptical-like ones. 
A more detailed analysis of completeness will 
be presented in a forthcoming paper (Cresci et al. \cite{cresci}).

\subsection{Confirming that we are accounting for anisoplanaticism}

\begin{figure}
    \begin{center}
    \resizebox{\hsize}{!}{\includegraphics{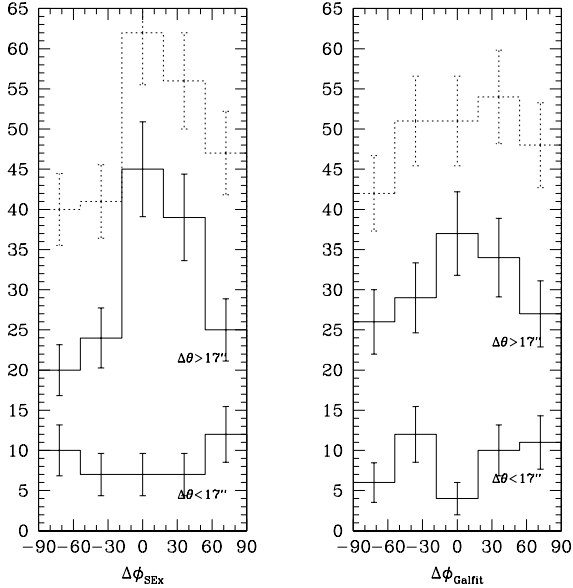}}   
    \caption{The difference $\Delta \phi$ between the radial vector 
    from the guide star to the source at offset ($\Delta x, \Delta y$) 
    and the source's own major axis 
    position angle $\phi$ as measured with SExtractor (\textit{left panel}) 
    with no PSF correction and with GALFIT (\textit{right panel}) using the model 
    PSF. In each panel the distribution for objects with angular distance from 
    the guide star $<17\arcsec$ and $\geq 17\arcsec$ are shown (solid lines) 
    as well as the total distribution of all sources (dashed line). 
    Error bars are $\sqrt{N}$ for $N$ the number of objects in each bin.}
    \label{posang}
    \end{center}
\end{figure}

The recovered position angle distribution can be used to indicate whether we are
indeed correcting for the anisoplanatic PSF in an effective manner,
and to evaluate possible biases 
in the results due to contributions by the wavefront correction that are still 
uncorrected by our PSF model.
We have divided our sample in two subsets with separations
$\Delta\theta<17\arcsec$ and $\Delta\theta \geq 17\arcsec$. 
This separation was chosen empirically as the approximate 
radius beyond which the PSF distortion becomes 
noticeable (see Baker et al. \cite{baker05}), but also corresponds 
to the average isoplanatic angle of the five fields in our analysis
here. 
If no radial stretch due to PSF anisoplanaticism is present (or if
this effect is fully taken into account by a PSF model) we would
expect the distribution of  
$\Delta\phi=\phi - \arctan(\Delta x/\Delta y)$, 
i.e., the difference between the radial vector from the star to the
source at offset ($\Delta x, \Delta y$) and the source's own major
axis position angle $\phi$, to be uniform over the interval
$[-90^\circ,+90^\circ]$ for both $\Delta\theta$ subsamples. 

The distribution derived using SExtractor with no PSF 
anisoplanaticism correction (see Fig.~\ref{posang}, left panel) shows
a clear peak at $\Delta\phi=0$ for objects at large radial distances
from the guide star, i.e., these objects appear to be preferentially
oriented towards the guide star.
The right panel of Fig.~\ref{posang} shows the same distribution as measured 
by GALFIT using the model PSF. 
As one would expect, in this case the peak is greatly reduced due to
the model PSF.
The small remaining overcounts could easily be due to random chance
rather than a systematic effect.
Indeed, analysis of this effect in the simulated galaxies indicates
that it is completely removed.

\section{Conclusions} \label{concl}

In this paper we have presented a new approach to account for the PSF 
variations across the field of view in deep AO images, developed to 
correct the anisoplanaticism in SWAN images obtained with NACO at the ESO VLT,
but also more generally applicable to other wide-field adaptive optics
data. 
The survey is intended to overcome the current shortage of extragalactic AO
targets and to study faint and compact field galaxies with unprecedented         
resolution in the near-IR. The observing strategy fully exploits the present 
capabilities of AO instrumentation, so that our method for PSF modeling 
may be broadly useful for AO cosmology.


We have described the PSF as the convolution between the 
on-axis PSF and a spatially varying kernel, an elliptical Gaussian elongated 
towards the AO guide star. We find that even adopting the most extreme case 
in which the the difference between the kernels in two different fields is 
given by a single parameter, namely the isoplanatic angle, the 
PSF can still be described with enough accuracy to extract reliable 
morphological parameters of field galaxies.  
This approach is particularly convenient, as it uses only easily available data 
and makes no uncertain assumptions about the stability of the isoplanatic angle 
during any given night. 
In addition, just a few point sources are required in each 
field to derive the isoplanatic angle and therefore compute the
kernel.

Our simulations demonstrate that the model is able to recover reliably 
morphological parameters in a typical SWAN field with an integration time of 
one hour up to $K_\mathrm{s} \sim 20.5$ 
($K_{\mathrm{AB}} \sim 22.3$), and that our 
very simple model is nearly as good as using the true PSF to account for 
anisoplanaticism.

Finally, we have presented the first morphological results for five SWAN fields, 
using the GALFIT package coupled with our model PSF. The recovered source
parameters confirm that we are indeed accounting for anisoplanaticism, 
and show the 
unique power of AO observations to derive the details of morphology in faint 
galaxies. These results pave the way for the forthcoming analysis of all the 
obtained SWAN fields, which will combine for the first time the high angular 
resolution of a space-based survey with the shallower depth and wider area of 
a ground-based survey.
  

\begin{acknowledgements}

The authors are grateful to the staff at Paranal Observatory for their
hospitality and support during the observations. We thank Rainer 
Sch\"odel for obtaining the observations of SBSF\,41, Nancy Ageorges and 
Chris Lidman for helping us sift through the NACO data archive; and our 
collaborators on SWAN (Reihnard Genzel, Reiner Hofmann, Sebastian Rabien, 
Niranjan Thatte, and W. Jimmy Viehhauser); Christophe Verinaud, Miska Le Louarn, 
Emiliano Diolaiti and Thierry Fusco for their assistance.
Some of the data included in this paper were obtained as part of the MPE
guaranteed time programme. GC and AJB acknowledge MPE for supporting 
their efforts on this project; AJB further acknowledges support from 
the National Radio Astronomy Observatory, which is operated by Associated 
Universities, Inc., under cooperative agreement with the National 
Science Foundation.
\end{acknowledgements}


\end{document}